\documentclass[aps,prl,twocolumn,nofootinbib,superscriptaddress,showpacs,floatfix,preprintnumbers]{revtex4-2}

\usepackage[dvipsnames]{xcolor}     
\definecolor{lcolor}{rgb}{0.5,0,0}
\definecolor{citcolor}{rgb}{0,0,1}
\usepackage[breaklinks,colorlinks,urlcolor=blue,citecolor=citcolor,linkcolor=lcolor,linktoc=all]{hyperref}
\usepackage{color}
\usepackage{graphicx}	
\graphicspath{{./figures/}}
\usepackage[utf8]{inputenc}
\usepackage{amsmath} \usepackage{amssymb}
\usepackage[dvipsnames]{xcolor}      
\usepackage{bm,bbm,bbold}
\usepackage{booktabs}
\usepackage{dcolumn}

\makeatletter
\def\@fnsymbol#1{\ensuremath{\ifcase#1\or *\or \dagger\or \ddagger\or
   \mathsection\or \mathparagraph\or \|\or **\or \dagger\dagger
   \or \ddagger\ddagger \else\@ctrerr\fi}}
\makeatother

\allowdisplaybreaks

\makeatletter
\g@addto@macro\bfseries{\boldmath}
\makeatother

\usepackage{tikz}
\usepackage[customcolors]{hf-tikz}
\usepackage{mciteplus}

\makeatletter 

\renewcommand\onecolumngrid{
\do@columngrid{one}{\@ne}%
\def\set@footnotewidth{\onecolumngrid}
\def\footnoterule{\kern-6pt\hrule width 1.5in\kern6pt}%
}

\renewcommand\twocolumngrid{
        \def\footnoterule{
        \dimen@\skip\footins\divide\dimen@\thr@@
        \kern-\dimen@\hrule width.5in\kern\dimen@}
        \do@columngrid{mlt}{\tw@}
}%

\makeatother    

\newcommand{\dd}{\mathrm{d}}

\newcommand{\mt}[1]{\textrm{\scriptsize #1}}
\def\Nc{N_\mt{c}}
\def\Nf{N_\mt{f}}
\def\Vf{V_\mt{f}}
\def\Vg{V_\mt{g}}
\def\rh{r_\mt{h}}
\def\Mp{M_\mt{p}}
\def\nB{n_\mt{B}}
\def\nq{n_\mt{q}}

\newcommand{\vev}[1]{{\left\langle #1 \right\rangle}}

\begin{document}

\title{Estimate for the bulk viscosity of strongly coupled quark matter {\\ \textcolor{black}{using perturbative QCD and holography}}}

\preprint{APCTP Pre2024 - 003}
\preprint{HIP-2024-1/TH}
\preprint{TUM-EFT 187/24}

\author{Jes\'us Cruz Rojas}
\email{jesus.cruz@correo.nucleares.unam.mx}
\affiliation{Asia Pacific Center for Theoretical Physics, Pohang, 37673, Korea}
\affiliation{Departamento de F\'isica de Altas Energ\'ias, Instituto de Ciencias Nucleares,
Universidad Nacional Aut\'onoma de M\'exico,
Apartado Postal 70-543, CDMX 04510, M\'exico}
\author{Tyler Gorda}
\email{gorda@itp.uni-frankfurt.de}
\affiliation{Institut f\"ur Theoretische Physik, Goethe Universit\"at,   Max-von-Laue-Str.~1, 60438 Frankfurt am Main, Germany}
\affiliation{Technische Universit\"{a}t Darmstadt, Department of Physics, 64289 Darmstadt, Germany}
\affiliation{ExtreMe Matter Institute EMMI, GSI Helmholtzzentrum f\"ur Schwerionenforschung GmbH, 64291 Darmstadt, Germany}
\author{Carlos Hoyos}
\email{hoyoscarlos@uniovi.es}
\affiliation{Departamento de F\'{\i}sica and Instituto de Ciencias y Tecnolog\'{\i}as Espaciales de Asturias (ICTEA), Universidad de Oviedo,c/ Leopoldo Calvo Sotelo 18, ES-33007, Oviedo, Spain}
\author{Niko Jokela}
\email{niko.jokela@helsinki.fi}
\affiliation{Department of Physics and Helsinki Institute of Physics,
P.O.~Box 64, FI-00014 University of Helsinki, Finland}
\author{Matti J\"arvinen}
\email{matti.jarvinen@apctp.org}
\affiliation{Asia Pacific Center for Theoretical Physics, Pohang, 37673, Korea}
\affiliation{Department of Physics, Pohang University of Science and Technology, Pohang, 37673, Korea}
\author{Aleksi Kurkela}
\email{aleksi.kurkela@uis.no}
\affiliation{Faculty of Science and Technology, University of Stavanger, 4036 Stavanger, Norway}
\author{Risto Paatelainen}
\email{risto.paatelainen@helsinki.fi}
\affiliation{Department of Physics and Helsinki Institute of Physics,
P.O.~Box 64, FI-00014 University of Helsinki, Finland}
\author{Saga S\"appi}
\email{saga.saeppi@tum.de}
\affiliation{Technical University of Munich
TUM School of Natural Sciences
Department of Physics, James-Franck-Str.~1, 85748 Garching, Germany}
\affiliation{Excellence Cluster ORIGINS, Boltzmannstrasse 2, 85748 Garching, Germany}
\author{Aleksi Vuorinen}
\email{aleksi.vuorinen@helsinki.fi}
\affiliation{Department of Physics and Helsinki Institute of Physics,
P.O.~Box 64, FI-00014 University of Helsinki, Finland}

\begin{abstract}
Modern hydrodynamic simulations of core-collapse supernovae and neutron-star mergers require knowledge not only of the equilibrium properties of strongly interacting matter, but also of the system's response to perturbations, encoded in various transport coefficients. 
Using perturbative and holographic tools, we derive here an improved weak-coupling and a new strong-coupling result for the most important transport coefficient of unpaired quark matter, its bulk viscosity. 
These results are combined in a simple analytic pocket formula for the quantity that is rooted in perturbative Quantum Chromodynamics at high densities but takes into account nonperturbative holographic input at neutron-star densities, where the system is strongly coupled. This expression can be used in the modeling of unpaired quark matter at astrophysically relevant temperatures and densities.
\end{abstract}

\maketitle

\section{Introduction}

During the last ten years, neutron stars (NSs) and their binary mergers --- observable through both electromagnetic and gravitational waves (GW) \cite{LIGOScientific:2017vwq,LIGOScientific:2017ync} --- have established themselves as the leading laboratory for dense Quantum Chromodynamics (QCD) matter. 
While the observable properties of single quiescent NSs and even the inspiral parts of NS mergers are mostly determined by the equation of state (EoS) of the constituent matter, the ringdown phase of a NS merger constitutes a considerably more complicated out-of-equilibrium system. 
In preparation for the eventual observation of a ringdown GW signal, extensive hydrodynamic simulations of NS mergers are currently being carried out, with one crucial challenge being to correctly account for energy dissipation and transport in {\color{black} NS matter} \cite{Baiotti:2016qnr}. 

Among the different transport coefficients, the bulk viscosity $\zeta$, which quantifies energy dissipation during a rapid compression or expansion of matter, stands out as particularly  important~\cite{Sawyer:1989dp,Haensel:1992zz,Haensel:2000vz,Alford:2018lhf,Alford:2019qtm,Alford:2019kdw,Alford:2021ogv,Alford:2023gxq,Yang:2023ogo}. 
For isolated NSs, it affects the emission of continuous GWs \cite{Sieniawska:2019hmd}, expected to be detectable in next-generation GW observatories such as the Einstein Telescope \cite{Punturo:2010zz} and Cosmic Explorer \cite{Reitze:2019iox}, and determines the maximal rotation frequencies of pulsars in a temperature-dependent fashion, giving rise to the so-called $r$-mode stability window {\color{black} in the 1-100 keV range}~\cite{Alford:2010fd,Alford:2013pma,Alford:2015gna} (for a review of NS oscillatory modes, see \cite{Kruger:2021zta}). 
In NS mergers, the bulk viscosity on the other hand provides damping for density oscillations, affecting both the inspiral \cite{Ripley:2023lsq} and post-merger dynamics,  {\color{black} of which the latter involves temperatures up to tens of MeVs. 
The bulk viscosity} may indeed leave a detectable imprint on the post-merger GW waveform \cite{Alford:2017rxf,Hammond:2022uua,Most:2022yhe,Alford:2022ufz,Chabanov:2023blf,Chabanov:2023abq}, the magnitude of which is however still under discussion \cite{Radice:2021jtw}.

The dominant contribution to the bulk viscosity comes about when weak interactions cannot keep pace with the compression rate, leading to deviations from beta equilibrium and a non-equilibrium contribution to the pressure, against which work can be done.
This effect peaks when the timescales of macroscopic oscillations and microscopic flavor-changing rates match.
In the nuclear matter phase, the value of $\zeta$ depends on multiple factors, such as whether direct Urca processes are allowed or if hyperons or Cooper pairing between nucleons are present, each affecting in particular the temperature scale where $\zeta$ reaches its maximal value (see ref.~\cite{Schmitt:2017efp} for a review).

\begin{figure}
    \includegraphics[width=0.48\textwidth]{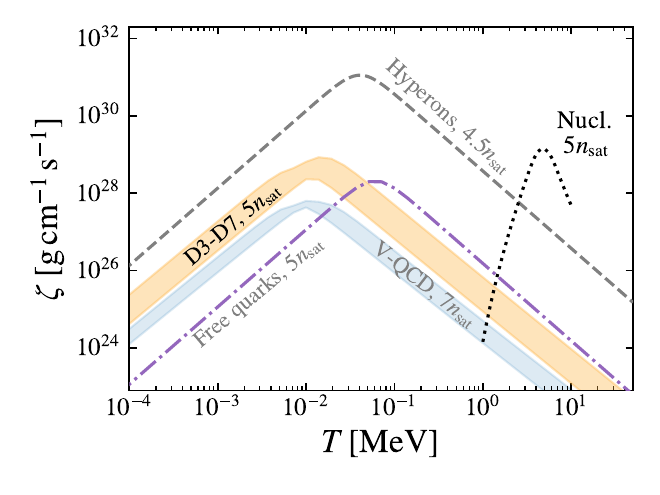}
    \caption{{\color{black} The bulk viscosity $\zeta$ of NS matter, evaluated at rotation frequency $\omega = 2 \pi \times 1$~kHz and given as a function of  $T$ for a baryon density $n_\mathrm{B}\approx 5n_\mathrm{sat}$.
    The uncertainty bands of the holographic results are assessed via their matching to QCD: The D3-D7 result is matched to pQCD quark densities 
including their uncertainty bands, while the uncertainty of the V-QCD result is estimated by varying the parameters of the model within limits set by  lattice-QCD results. Finally, nuclear and hyperonic matter results (labeled ``Nucl.'' and ``Hyperons'') from Refs.~\cite{Alford:2022ufz,Jyothilakshmi:2022hys} are shown for comparison. Note that for technical reasons, the V-QCD result is shown for $7n_\mathrm{sat}$ and the hyperonic one for $4.5n_\mathrm{sat}$. We observe that {\textcolor{black}{our}} QM results always peak within the $r$-mode stability window 1-100 keV, but are strongly suppressed at the $O(10~\text{MeV})$ temperatures involved in NS mergers. {\textcolor{black}{This may, however, be related to the absence of quark pairing in our setup (see \cite{Alford:2008pb} for a counterexample in the Color-Flavor-Locked case).}}}
    \label{fig:zeta_func_T_approx}}
\end{figure}

The first milliseconds of a binary NS merger are known to involve baryon densities up to several nuclear saturation densities $n_\mathrm{sat}\approx 0.16/\mathrm{fm}^3$ as well as temperatures up to several tens of MeV's (see, e.g., \cite{Hanauske:2016gia}).
Such conditions may also lead to the creation of deconfined QM \cite{Annala:2019puf,Annala:2021gom,Annala:2023cwx,Prakash:2021wpz,Tootle:2022pvd}, the transport properties of which differ significantly from those of nuclear matter.
While the value of the QM bulk viscosity is expected to strongly depend on the presence and details of quark pairing, differences between various partially paired configurations are expected to be smaller than between quark and nuclear matter \cite{Schmitt:2017efp}. 
This makes the bulk viscosity an interesting quantity for tracking the possible creation of QM during mergers.

Despite the phenomenological importance of the bulk viscosity, our ability to predict its behavior remains limited owing to the unavailability of controlled first-principles quantum-field-theory methods at NS densities. 
The leading first-principles tools include perturbative QCD (pQCD), available only at very high densities {\color{black}  (see, e.g., \cite{Kurkela:2009gj, Gorda:2023usm, Gorda:2023mkk})}, and holography, which describes the strong-coupling limit of a class of QCD-like theories \cite{Casalderrey-Solana:2011dxg,Brambilla:2014jmp,Jarvinen:2021jbd,Hoyos:2021uff,Rougemont:2023gfz}. 
For QM, leading-order perturbative results for several transport coefficients were derived some thirty years ago \cite{Madsen:1992sx,Heiselberg:1993cr} and improved to next-to-leading order (NLO) later \cite{Sad:2006egl,Sad:2007afd}, whereas at strong coupling, the shear viscosity and the electrical and thermal conductivities were first evaluated only recently in two holographic models \cite{Hoyos:2020hmq,Hoyos:2021njg}. 
{\color{black}For the bulk viscosity, only the minuscule purely QCD contribution has been considered in recent literature \cite{Czajka:2018bod,Hoyos:2020hmq}, but for the dominant contribution stemming from an interplay between the electroweak and strong sectors, no strong-coupling prediction is currently available at all.}

In this work, we derive state-of-the-art results for the thermodynamic response of QM to a change in its flavor content, thus providing novel predictions for the bulk viscosity. 
We do so using both perturbative and holographic methods, and in particular derive the first strong-coupling predictions for the quantity.
Our results are applicable for unpaired QM and serve as a starting point for any partially unpaired phase \cite{Schmitt:2017efp}.

{\color{black}The main result of our work is shown in fig.~\ref{fig:zeta_func_T_approx}, where we display the bulk viscosity of NS matter as a function of temperature for a baryon density of roughly $5n_\mathrm{sat}$. 
For QM, we include results corresponding to the free-theory limit,  evaluated at a fixed strange quark mass ($m_s = 93.4$~MeV), as well as our two holographic models, D3-D7 and V-QCD, but not the pQCD result, which is not under quantitative control at intermediate densities. 
For the confined phase, we display results corresponding to both nuclear \cite{Alford:2022ufz} and hyperonic \cite{Jyothilakshmi:2022hys} matter.
As we discuss in detail in the remaining sections of this letter, our results paint a consistent picture of the behavior of the QM bulk viscosity that displays a stark qualitative difference to that witnessed in the confined phases of QCD.}
Furthermore, we observe that for astrophysically relevant densities and temperatures, nearly all temperature dependence in the QM result originates from the flavor-changing interactions. 
For our D3-D7 computation, this leads to a simple analytic result for $\zeta$, given in eq.~(\ref{eq:zeta2}) below, that we suggest for use as an approximation for the bulk viscosity of unpaired QM in future phenomenological applications.

\section{Setup}

For unpaired {\color{black}three-flavor} QM in the neutrino-transparent regime, the leading contribution to the bulk viscosity arises from $W$-boson exchange in the process $u+d\longleftrightarrow u+s$.
Outside beta equilibrium, i.e., when the $d$ and $s$ quark chemical potentials differ $\mu_d\neq \mu_s$,  the quark densities $n_d$ and $n_s$ change with rates proportional to an electroweak rate $\lambda_1$ \cite{Heiselberg:1986pg,Heiselberg:1992bd,Madsen:1993xx}, so that
\begin{equation}
    \frac{\dd n_d}{\dd t}=-\frac{\dd n_s}{\dd t}\approx \lambda_1(\mu_s-\mu_d) \ .
\end{equation}
Neglecting quark masses, the leading low-$T$ contribution to the rate becomes \cite{Madsen:1993xx}\footnote{We have checked that the free-quark $\zeta$ following from eq.~\eqref{eq:lambda1} agrees with the unpaired results of \cite{Alford:2006gy}, where the numerical rate was evaluated without the small-$T$ or small-$m_s$ approximations.}
\begin{eqnarray}\label{eq:lambda1}
    \lambda_1 &=& {\color{black}\bigg(1+\sigma \log\frac{\Lambda}{T}\bigg)^4}\frac{64}{5\pi^3} G_F^2 \sin^2 \theta_c \cos^2 \theta_c \mu_d^5 T^2 \, ,
\end{eqnarray}
{\color{black}where $G_F$ is the Fermi constant and $\theta_c$ the Cabibbo angle. The quartic prefactor on the right-hand side represents the only known $O(\alpha_s)$ correction to the rate, which is moreover logarithmically enhanced at low temperatures as it originates from a so-called non-Fermi-liquid (nFL) contribution to the specific heat of QM \cite{Schwenzer:2012ga} (see also \cite{Gerhold:2004tb,Schafer:2004zf}). 
As discussed in detail around fig.~\ref{fig:nFLeffect} of the Supplemental Material, this correction allows us to gauge the importance of the (partially unknown) $\mathcal{O}(\alpha_s)$ corrections to the rate: for $\sigma=0$, the result reduces to the leading-order rate, while for $\sigma\equiv 4\alpha_s/(9\pi)$ and $\Lambda\approx 0.158\sqrt{\alpha_s}\sqrt{\mu_u^2 + \mu_d^2 + \mu_s^2}$ one recovers the result derived in \cite{Schwenzer:2012ga}.}

{\color{black}While the unknown QCD corrections to the rate may be sizable, we note that the qualitative behavior of the rate likely remains the same at strong coupling:} 
In holography, the QCD contribution to the rate, replacing the leading-order multiplicative factor $\mu_d^5 T^2$ above, is available from the convolution of two flavor-current correlators. 
For these correlators, calculations at non-zero quark densities in the D3-D7 model show a linear dependence on the temperature at low frequencies \cite{Erdmenger:2007ja,Mas:2008jz,Erdmenger:2008yj}, consistent with the formula we use. 
Furthermore, the normalization of the correlators depends on the number of colors and flavors but not on the 't Hooft coupling, thus keeping the rate constant in the strong-coupling limit.

A study of energy dissipation during a compression-decompression cycle near beta equilibrium connects $\zeta$ to various susceptibilities $\chi_{ij} \equiv \partial^2 p / \partial \mu_i \partial \mu_j$ and reaction rates (see Supplemental Material sec.~A).\footnote{The breaking of $\beta$ equilibrium in mergers is discussed in \cite{Alford:2018lhf, Yang:2023ogo}.} 
If we only take into account the $u+d\longleftrightarrow u+s$ process~\cite{Sad:2007afd}, this leads to 
\begin{equation}\label{eq:zeta}
    \zeta = \frac{\lambda_1 A_1^2}{\omega^2+(\lambda_1 C_1)^2} \ ,
\end{equation}
{\color{black}where the coefficients $A_1$ and $C_1$, determined by various susceptibilities and quark densities, are found in eqs.~(\ref{A1}) and (\ref{C1}) of the Supplemental Material and} $\omega$ denotes the angular frequency of density oscillations (see \cite{Kruger:2021zta} for discussion).\footnote{To express the frequency in the high-energy physics units of $\mathrm{MeV}$ instead of $\mathrm{Hz}$, we use $ \mathrm{MeV}~\equiv 4.1351 \times 10^{-21} \mathrm{rad/s}~\equiv 2\pi \times 6.58122\times 10^{-22} \mathrm{Hz}$. Similarly, should one further wish to express $\zeta$ in astrophysical units, the relevant conversion factor from the high-energy-physics units is $ \mathrm{MeV}^3 = 137 286 \times \mathrm{g \,cm^{-1}\, s^{-1}}$.} 

The combination of susceptibilities appearing in eq.~(\ref{A1}) vanishes if the $d$ and $s$ quarks are degenerate in mass --- a fact most easily verified if  (\ref{A1}) is given in terms of the inverse susceptibility matrix (see Supplemental Material for details). 
This implies that a nonzero strange quark mass must be implemented in both the weak- and strong-coupling setups, which we briefly introduce below.

\section{Methods}

In this section, we review our perturbative and holographic determinations of the susceptibilities that enter eq.~(\ref{eq:zeta}). 
In both calculations, we treat electrons as non-interacting and (numerically) solve the corresponding chemical potential $\mu_e$ from the charge neutrality condition $2 n_u/3 -  n_d/3 - n_s/3 = n_e = T^2\mu_e/3 -\mu_e^3/(3\pi^2)$. 
Together with the beta-equilibrium conditions $\mu_s=\mu_d,$ $\mu_u = \mu_d - \mu_e$, this allows us to obtain $\zeta$ in terms of $\mu_d$, $T$, $\omega$. Finally, our results will depend on the parameter $X \equiv \bar{\Lambda}/(2\mu_d)$ which parametrizes our results' dependence on the unphysical renormalization scale $\bar{\Lambda}$ in the $\overline{\mathrm{MS}}$ scheme. It appears directly in our pQCD results and indirectly in the D3-D7 ones, where it enters through the high-density matching of the model to pQCD. 

\subsection{Perturbative QCD}

For vanishing quark masses, the perturbative pressure of deconfined unpaired QCD matter is known up to order $\alpha_s^{5/2}$ at nonzero temperatures and densities \cite{Vuorinen:2003fs,Kurkela:2016was} and up to partial $\mathcal{O}(\alpha_s^{3})$ in the $T=0$ limit \cite{Gorda:2018gpy,Gorda:2021znl,Gorda:2023mkk}.
Up to the highest fully known order $\mathcal{O}(\alpha_s^{5/2})$, the result can be split into two distinct terms corresponding to contributions from the hard and soft momentum scales, which for $\mu\gg T$ are of order $\mu$ and $\alpha_s^{1/2} \mu$, respectively.
We treat the additional mass-dependent contribution to the pressure $p_m$ within the mass-expansion scheme of \cite{Gorda:2021gha}, where $m_s$ is formally treated as a quantity of $\mathcal{O}(\alpha_s^{1/2} \mu)$ and the light quark masses are neglected.
This mass expansion is performed to $\mathcal{O}(m_s^4)$ and up to a combined $\mathcal{O}(\alpha_s^{5/2})$ (for the full mass-dependence at $T=0$, see \cite{Kurkela:2009gj}).
For the value of the $s$ quark mass, we use the physical $\overline{\mathrm{MS}}$ renormalized value $m_s \approx 93.4\, \text{MeV}$  \cite{Workman:2022ynf}. 
We have confirmed that additionally including nonzero $m_u$ and $m_d$ terms would lead to a vanishingly small effect,
\textcolor{black}{while the chemical potentials realized in NSs are not large enough to allow for heavier quarks.}
For the soft contribution, evaluated in the massless limit, we furthermore use an analytic small-$T/\mu$ expansion  derived in \cite{Kurkela:2016was} that is valid for $T\lesssim 100$~MeV. Mass corrections to this result start at $\mathcal{O}(\alpha_s^3)$ and can therefore be neglected. 

The perturbative pressure described above can be readily differentiated to obtain predictions for the coefficients $A_1,C_1$ and eventually for $\zeta$ as functions of the three quark chemical potentials and the renormalization scale parameter $X$. 
The results constitute lengthy closed-form expressions in terms of standard special functions and their derivatives, allowing for inexpensive evaluation of the necessary quantities. 

\subsection{Holography}

The D3-D7 model \cite{Karch:2002sh} is the holographic dual of ${\cal N}=4$ $\mathrm{SU}(\Nc)$ super Yang--Mills theory with $\Nf$ copies of ${\cal N}=2$ hypermultiplets in the quenched approximation $\Nf/\Nc\ll 1$. 
It consists of $\Nf$ probe  D7-branes embedded in the $\mathrm{AdS}_5\times S^5$ spacetime, while baryon charge is introduced by turning on an electric field on the D7-branes \cite{Kobayashi:2006sb,Mateos:2007vc} and temperature by modifying the geometry to that of a black brane. 
Following \cite{Hoyos:2016zke}, we extrapolate the model to the physically relevant $\Nc=\Nf=3$ and fix {\color{black}$\alpha_s\approx 0.285$} so that the pressure matches the Stefan--Boltzmann value at high density, extending the model's validity towards higher densities. 
{\color{black} Although the field content of the model differs from that of QCD, we note that the thermodynamic coefficients $A_1$ and $C_1$, obtained through chemical-potential derivatives of the pressure, are highly insensitive the additional fields in the D3-D7 model.}

At vanishing temperature, the pressure of the D3-D7 model takes the simple form \cite{Karch:2007br,Hoyos:2016zke}
\begin{equation}
\label{eq:d3d7pres}
    p=\frac{1}{4\pi^2}\sum_{i=u,d,s}(\mu_i^2-M_i^2)^2 \ ,
\end{equation}
where $M_i$ are the constituent quark masses that we fix by equating quark densities with pQCD at $\mu_d=1\,\text{GeV}$ and varying $X \in [1/2,2]$. 
Doing so, we obtain $M_u \in  (522.5,434.6) \,\text{MeV}$, $M_d \in  (526.4,435.9) \,\text{MeV}$, and $M_s \in (541.8,450.1) \,\text{MeV}$, within this interval in $X$. 
In what follows, in addition to estimating uncertainties by matching to pQCD at different values of $X$, we also vary this matching density within $\mu_d \in [1, 2]$~GeV. 
At $T\neq 0$, we finally compute the pressure numerically, following methods introduced in \cite{Kobayashi:2006sb,Mateos:2007vc}.

The other holographic model we use is V-QCD \cite{Jarvinen:2011qe}, which is a bottom-up model tuned to reproduce QCD physics as closely as possible (see, e.g., the reviews \cite{Gursoy:2010fj,Jarvinen:2021jbd,Hoyos:2021uff}). 
It combines the improved holographic QCD model for pure Yang--Mills theory~\cite{Gursoy:2007cb,Gursoy:2007er} to a description of flavors introduced via tachyonic brane actions~\cite{Bigazzi:2005md,Casero:2007ae,Bergman:2007pm}, {\color{black}featuring, e.g., a running $\alpha_s$ as reviewed in the Supplemental Material.} 
Given that quarks are treated as unquenched ($\Nf/\Nc\sim 1$) in V-QCD, the model should capture their physics more realistically than the D3-D7 model. 
Indeed, V-QCD by construction agrees with various qualitative properties of QCD (such as confinement and asymptotic freedom), and its parameters are fitted to data, including lattice results for the pressure \cite{Gursoy:2009jd,Jokela:2018ers} and baryon number susceptibilities \cite{Jokela:2018ers} at $\mu=0$. 
The model is consistent with all known astrophysical observations in the NS-matter regime~\cite{Jokela:2021vwy,Demircik:2021zll}, but eventually becomes inconsistent with pQCD at high densities \cite{Komoltsev:2021jzg}.

In this paper, we otherwise follow the treatment of the above V-QCD papers but relax the assumption of exact chiral symmetry in the QM phase by turning on a nonzero strange quark mass, thus extending the prescription of  \cite{Jarvinen:2015ofa}. 
The corresponding mass parameter of the model is fixed by demanding that the masses of kaons and $\eta$ mesons are well reproduced in the vacuum (see Supplemental Material {\textcolor{black}{and refs.~\cite{Amorim:2021gat,Jarvinen:2022gcc,Arean:2012mq,Arean:2013tja}}} for details). We find that this procedure under-predicts the dependencies of quark number susceptibilities on the strange quark mass at zero $\mu$ and high $T$, where the results can be benchmarked against lattice data~\cite{Borsanyi:2011sw}. This leads us to expect that this model similarly under-predicts the effects of the strange quark mass in physical quantities at high densities.

Finally, we quantify the underlying uncertainty of our results by allowing the V-QCD parameters vary within limits set by the lattice QCD fit in the chirally symmetric phase \cite{Jokela:2018ers,Ishii:2019gta}, but otherwise follow the computational strategy of \cite{Alho:2013hsa} in determining the quantities appearing in eq.~(\ref{eq:zeta}). 
In both holographic setups, the variation procedure we perform thus corresponds to
the uncertainties associated with the respective matching procedures.

\section{Results}

Our main result for the bulk viscosity of unpaired QM is displayed in  fig.~\ref{fig:zeta_func_T_approx}. It highlights a qualitative contrast between the behavior of $\zeta$ in {\color{black} the confined and deconfined phases of QCD, with the more suppressed QM results peaking at lower temperatures,} and in addition demonstrates the important effect of interaction corrections in the latter case. Consistently with our expectations for quantities that vanish in the degenerate-mass limit, V-QCD appears to predict somewhat lower values for $\zeta$ than our other methods, but nevertheless retains the same qualitative features. 

A closer inspection of our results reveals a number of interesting further findings. Explicit calculations show that in all three approaches, the bulk viscosity is insensitive to the $T$-dependence originating from {\color{black}the coefficients $A_1$ and $C_1$ of eq.~\eqref{eq:zeta}.} As demonstrated in fig.~\ref{fig:d3d7_exact_approx} of Supplemental Material, to a good accuracy we can indeed set $T=0$ in these functions and only keep the $T$-dependence of the electroweak rate $\lambda_1$ in eq.~\eqref{eq:lambda1}. Another universal characteristic that all our results exhibit is an approximate quartic dependence on the strange quark mass which has been noted before in \cite{Alford:2006gy}.

\begin{figure}
    \includegraphics[width=0.48\textwidth]{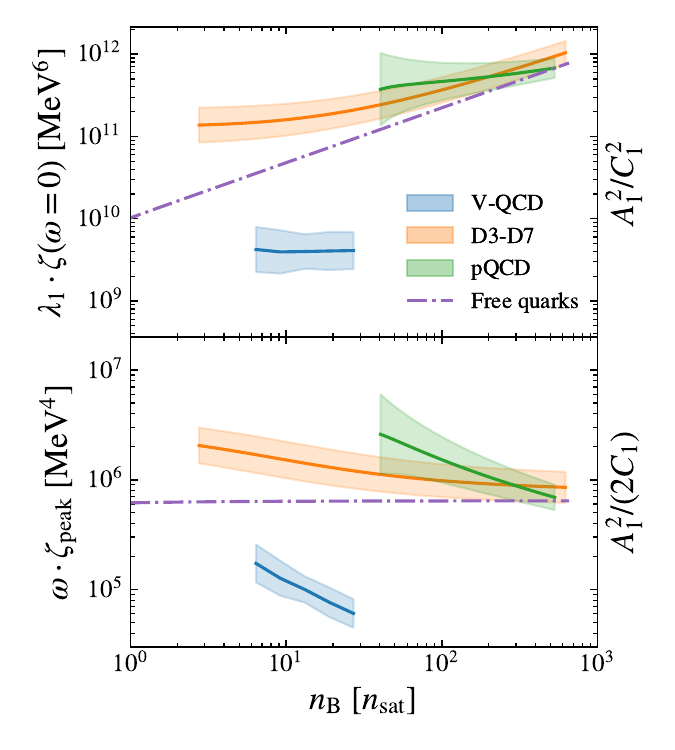}
    \caption{A comparison of the values of two quantities characterizing the bulk viscosity: its zero-frequency limit and peak value, $\zeta (\omega = 0)$ and $\zeta_\mathrm{peak}$. 
    These quantities are multiplied by different factors so that they depend only on $A_1$ and $C_1$ in eqs.~\eqref{A1} and \eqref{C1} and are independent of the oscillation frequency $\omega$ and the electroweak rate $\lambda_1$, as indicated by the expressions on the right vertical axes.
    The error bars in these panels capture the variation of model parameters in the different models as described in the main text. 
    }
    \label{fig:zeta_breakdown_3}
\end{figure}

While the full $\zeta$ depends on the rate $\lambda_1$, we may construct physical features of the bulk viscosity that are sensitive only to QCD input. 
{\color{black}For example, the peak value of the viscosity, $\zeta_{\rm peak} \equiv \zeta(T_{\rm peak})$, and its rescaled zero-frequency limit $\lambda_1 \zeta(\omega=0)$ that corresponds to the DC bulk viscosity entering the Israel--Stewart theory \cite{Israel:1979wp, Gavassino:2020kwo, Yang:2023ogo}} are completely insensitive to the electroweak rate and can be fully extracted from the coefficients $A_1$ and $C_1$ in eqs.~\eqref{A1}--\eqref{C1}. 
These two quantities are shown in fig.~\ref{fig:zeta_breakdown_3}, where 
we observe a good agreement between our pQCD and D3-D7 results for densities where both predictions are available, {\color{black}while V-QCD again appears to underestimate the quantities (see discussion in Supplemental Material).}

Setting $T=0$ in $A_1$ and $C_1$, we find that the D3-D7 calculation leads to a remarkably simple analytic formula as a function of $\mu_d$
\begin{equation}\label{eq:zeta2}
    \zeta= \,\frac{ 4\lambda _1 \, \mu _d^6\,(M_s^2-M_d^2)^2}{K_d^2 K_s^2\omega ^2 +\pi ^4
   \lambda _1^2\, (K_d+K_s)^2} \ ,
\end{equation}
where we have defined $K_i\equiv3\mu_d^2-M_i^2$.
We stress that for the $M_i$ in this formula, one should use the constituent quark-mass ranges listed below eq.~\eqref{eq:d3d7pres}, leading to the uncertainty ranges visible in fig.~\ref{fig:zeta_func_T_approx}.
To express this as function of $n_\mathrm{B}$ for the small temperatures of relevance to BNS mergers, one can further use the $T = 0$ pressure in eq.~\eqref{eq:d3d7pres} to numerically relate $n_\mathrm{B}$ to $\mu_d$ in beta equilibrium.

Returning finally to the bulk viscosity itself, we note that it is straightforward to compare our NNLO pQCD results to lower perturbative orders, as shown in fig.~\ref{fig:nlovsnnlo} of the Supplemental Material. 
We find that the difference between the NLO and NNLO results is non-negligible even at $40n_\mathrm{sat}$, and that the results diverge rapidly at lower densities, making extrapolation to the NS realm impossible.
While the naive free quark expression can, in principle, be extrapolated to low densities, it completely fails to take into account the effects of interactions, which become increasingly important much before the hadronic phase is eventually reached. 
For phenomenological purposes, the compact D3-D7 bulk viscosity of eq.~(\ref{eq:zeta2}) is on the other hand appealing as it is rooted in pQCD but takes into account the strongly coupled nature of the theory at low densities.
To this end, {\color{black} despite its limitations discussed above,} we recommend the use of this result in the modeling of dense unpaired QM at astrophysically relevant densities and temperatures, and similarly expect the the V-QCD result to provide a reasonable lower bound for the bulk viscosity.

An important limitation of our present approach is finally related to the fact that the pairing channel and the magnitude of the superconducting gap in low- and moderate-density QM remains unknown (though see \cite{Kurkela:2024xfh} for a recent model-independent study bounding the gap at high densities). 
To obtain estimates for the bulk viscosity in various pairing channels, corrections to both the electroweak rate in eq.~\eqref{eq:lambda1} and to the thermodynamic functions entering through eqs.~\eqref{A1}--\eqref{C1} should be separately considered. 
While the latter are expected to be subleading, the former may be substantial given that the contribution of gapped quark modes to the reaction rate is exponentially suppressed. 
{\color{black}While the detailed evaluation of these corrections is left for future work, we note that the electroweak rate receives $O(\alpha_s)$ QCD corrections even in the unpaired phase, some of which are presently known \cite{Schwenzer:2012ga}. Their effect is studied in fig.~\ref{fig:nFLeffect} of the Supplemental Material, where we observe that, in agreement with the $\lambda_1$-independence of $\zeta_\mathrm{peak}$, they primarily simply shift the peak of the viscosity to lower temperatures.}

\emph{Acknowledgements.}---%
We thank Arus Harutyunyan for providing us with the data for the nuclear bulk viscosities {\color{black} and V. Sreekanth for the hyperonic bulk viscosities, both} included in our fig.~\ref{fig:zeta_func_T_approx}, and we thank Mark Alford, Christian Ecker, Carlo Musolino, and Andreas Schmitt for helpful comments and suggestions on the manuscript.
J.C.R.~and M.J.~have been supported by an appointment to the JRG Program at the APCTP through the Science and Technology Promotion Fund and Lottery Fund of the Korean Government and by the Korean Local Governments --- Gyeong\-sang\-buk-do Province and Pohang City --- and by the National Research Foundation of Korea (NRF) funded by the Korean government (MSIT) (grant number 2021R1A2C1010834). 
T.G.~has been supported in part by the Deutsche Forschungsgemeinschaft (DFG, German Research Foundation) project-ID 279384907--SFB 1245, by the State of Hesse within the Research Cluster ELEMENTS (projectID 500/10.006), and by the ERC Advanced Grant ``JETSET: Launching, propagation and emission of relativistic jets from binary mergers and across mass scales'' (Grant No.~884631). 
C.H.~is partially supported by the AEI and the MCIU through the Spanish grant PID2021-123021NB-I00 and by FICYT through the Asturian grant SV-PA-21-AYUD/2021/52177. 
N.J., R.P., and A.V~have been supported in part by the Research Council of Finland grants no.~322507, 345070, 347499, 353772, and 354533 as well as by the ERC Consolidator grant no.~725369. 
S.S.~acknowledges support of the DFG cluster of excellence ORIGINS funded by the DFG under Germany’s Excellence Strategy - EXC-2094-390783311.

\bibliography{References.bib}

\onecolumngrid
\appendix
\newpage
\centerline{\bf \Large Supplemental Material}

\vspace{0.5cm}

In the three sections of the Supplemental Material, we go through several details of our calculations and results that provide additional context to the main text. 
These include a compact derivation of the main formulas for the bulk viscosity, eqs.~(\ref{eq:zeta})--(\ref{C1}), that we include here for completeness; details of how we have supplemented the V-QCD setup with a nonzero $s$ quark mass parameter; and a quantitative analysis and comparison of our three independent results for $\zeta$.

\section{A: In-equilibrium processes and bulk viscosity}

We review first the processes taking place in near-equilibrium unpaired quark matter and provide a relatively detailed derivation of the bulk viscosity formulas used in the main text. It follows the arguments introduced in \cite{Sawyer:1989dp,Haensel:2000vz} and reviewed in \cite{Schmitt:2017efp}, but presents more details and extends their analysis to include non-diagonal flavor susceptibilities.

\subsection{A.1: Beta equilibrium, charge neutrality, and production rates}

The relevant flavor-changing electroweak processes produced by the exchange of a $W$-boson are
\begin{equation}\label{eq:EWprocesses}
u+d\longleftrightarrow u+s\ , \qquad u+e \longleftrightarrow d+\nu_e\ ,\qquad u+e \longleftrightarrow s +\nu_e \ ,
\end{equation}
which lead to the chemical equilibration conditions (we assume no neutrino trapping, $\mu_{\nu_e}=0$)
\begin{equation}\label{eq:betaeq}
\mu_d=\mu_s\ ,\ \ \mu_u+\mu_e=\mu_d\ .
\end{equation}
In addition, local charge neutrality implies the following relation for the densities
\begin{equation}\label{eq:chargeneut}
n_Q=\frac{2}{3} n_u-\frac{1}{3}n_d-\frac{1}{3}n_s-n_e=0 \ .
\end{equation}

When the system is taken out of equilibrium, the densities of various particle species change according to \cite{Alford:2006gy}
\begin{equation}
\delta n_u=\delta n_e\ ,\ \ \delta n_d+\delta n_s=-\delta n_e\ 
\end{equation}
so that the baryon number and electric charge densities do not change,
\begin{equation}
\delta \nB=\frac{1}{3}\delta \nq=\frac{1}{3}\left(\delta n_u+\delta n_d+\delta n_s\right)=0\ ,\ \ \delta n_Q=0\ .
\end{equation}
For small deviations from beta equilibrium, the rates of change of the densities through the electroweak processes are proportional to
\begin{eqnarray}
\frac{\dd n_u}{\dd t} & = & \frac{\dd  n_e}{\dd t}\approx\lambda_2 (\mu_d-\mu_u-\mu_e)+\lambda_3(\mu_s-\mu_u-\mu_e) \nonumber\\
\frac{\dd  n_d}{\dd t} & \approx & \lambda_1(\mu_s-\mu_d)+\lambda_2(\mu_u+\mu_e-\mu_d) \\
\frac{\dd  n_s}{\dd t} & \approx & -\frac{\dd n_d}{\dd t}-\frac{\dd n_e}{\dd t}\ ,\nonumber
\end{eqnarray}
where $\lambda_i$ stand for the electroweak reaction rates for processes in eq.~(\ref{eq:EWprocesses}) enumerated left-to-right.

Since the baryon number and the electric charge are not affected by the above processes, we may use  the baryon or quark density and the following partial fractions as thermodynamic variables:
\begin{equation}\label{eq:partfrac}
X_a=\frac{n_a}{\nq},\ a=u,d\ ,\ \ X_s=1-X_u-X_d\ ,\ \ X_e=X_u-\frac{1}{3}\ .
\end{equation}
In this case, the reaction rates to consider are
\begin{equation}\label{eq:reacrates}
\begin{split}
&\nq\frac{\dd X_u}{\dd t}\approx \lambda_2 (\mu_d-\mu_u-\mu_e)+\lambda_3(\mu_s-\mu_u-\mu_e) \\
&\nq \frac{\dd  X_d}{\dd t}\approx \lambda_1(\mu_s-\mu_d)+\lambda_2(\mu_u+\mu_e-\mu_d)\ .
\end{split}
\end{equation}

\subsection{A.2: Formula for the bulk viscosity}

Changes in volume due to a radial pulsation of the system also change the baryon density. 
Considering the simple case of a small homogeneous oscillation of period $\tau=2\pi/\omega$, 
we can parameterize the deviation of the quark density $\nq$ from its equilibrium value $\nq^0$ as $\Delta n_q$, fulfilling
\begin{equation}
\nq\approx \nq^0+\Delta \nq \sin(\omega t)\ ,
\end{equation}
which can alternatively be seen as a change in the specific volume $V_\mt{q}=1/\nq$, such that
\begin{equation}\label{eq:dtVB}
\frac{\dd V_\mt{q}}{\dd t}\approx -\omega\frac{\Delta \nq}{(\nq^0)^2}\cos(\omega t)\ . 
\end{equation}

The oscillatory radial pulsation leads to the dissipation of energy $\dot{\cal E}$ due to a nonzero bulk viscosity of the medium, $\zeta$. 
Averaging over an oscillation period, we obtain
\begin{equation}\label{eq:dissen}
\vev{\dot{\cal E}_\text{diss}}=-\frac{\zeta}{\tau}\int_0^\tau \dd t (\nabla \cdot \bm{v})^2\ ,
\end{equation}
where $\bm{v}$ is the velocity of the fluid due to the pulsation. 
Using the continuity equation of baryon charge, we can further write the divergence of the velocity in the form
\begin{equation}
\dot{n}_\mt{q}+\nabla\cdot (\nq \bm{v})=0\ \ \Rightarrow \ \ \nabla \cdot \bm{v}=-\frac{\dot{n}_\mt{q}}{\nq}\approx -\omega\frac{\Delta \nq}{\nq^0}\cos(\omega t)\ ,
\end{equation}
so that the dissipated energy becomes
\begin{equation}
\vev{\dot{\cal E}_\text{diss}}=-\frac{\zeta \omega^2}{2} \left(\frac{\Delta \nq}{\nq^0} \right)^2\ .
\end{equation}
At the same time, the dissipated energy can be seen to result from the irreversibility of the compression-decompression cycle, for which the work performed in an infinitesimal change of the specific volume equals $d{\cal W}_q=p\, dV_\mt{q}$ with $p$ the pressure. 
Averaging over a period, we again obtain
\begin{equation}\label{eq:diss2}
\vev{\dot{\cal E}_\text{diss}}=\nq^0 \vev{\dot{\cal W}_\mt{q}} \approx \frac{\nq^0}{\tau}\int_0^\tau \, \dd t p(t) \frac{\dd V_\mt{q}}{\dd t}\ ,
\end{equation}
where a comparison with eq.~\eqref{eq:dtVB} shows that  nonzero contributions come from terms in $p(t)$ proportional to $\cos(\omega t)$. 

Let us next inspect the pressure, which is a function of chemical potentials $\mu_a$ in the grand canonical ensemble. 
A shift of these variables translates into
\begin{equation}
p=p_0+\sum_a \frac{\partial p}{\partial \mu_a} \delta \mu_a\ ,\ \ n_a=n_a^0+\sum_b \frac{\partial n_a}{\partial \mu_b} \delta \mu_b\ ,\ \ a,b=u,d,s,e\ ,
\end{equation}
where the partial derivatives are taken while leaving other thermodynamic variables fixed, namely the temperature and other chemical potentials. 
Denoting the matrix of susceptibilities as
\begin{equation}
\chi_a^{\ b}\equiv\frac{\partial n_a}{\partial \mu_b}\ ,
\end{equation}
it is then straightforward to derive the following relation between the rates from \eqref{eq:partfrac}:
\begin{equation}
\nq^0\frac{\dd X_a}{\dd t}=\chi_a^{\ b}\frac{\dd\mu_b}{\dd t}- X_a^0 \dot{n}_\mt{q}\ ,\ \ \dot{n}_\mt{q}=\omega \Delta \nq\cos(\omega t)\ .
\end{equation}
Using \eqref{eq:partfrac}, the pressure becomes in turn
\begin{equation}\label{eq:pressX}
p=p_0+ \nq^0\left[ X_u^0 (\delta \mu_u+\delta \mu_e-\delta \mu_s)+X_d^0 (\delta \mu_d-\delta \mu_s)+\delta \mu_s-\frac{1}{3}\delta \mu_e\right]\ .
\end{equation}

Next, we use known relations between the partial fractions of \eqref{eq:partfrac} to convert eq.~\eqref{eq:reacrates} into a set of differential equations for the chemical potentials. 
Defining the chemical potentials describing the deviation from the beta equilibrium
\begin{equation}
 \mu_1=\mu_s-\mu_d\ ,\ \ \mu_2=\mu_u+\mu_e-\mu_d\ ,
\end{equation}
we obtain ($\alpha,\beta=1,2$; $a=u,d,s,e$)
\begin{eqnarray}
\frac{\dd \mu_\alpha}{\dd t} & = & M_\alpha^a\left( \lambda_a^\beta \mu_\beta+X_a^0\dot{n}_\mt{q}\right)\nonumber\\
\frac{\dd \mu_s}{\dd t} & = & M_s^{\ a}\left(\lambda_a^\alpha\mu_\alpha+X_a^0 \dot{n}_\mt{q}\right)\\
\frac{\dd \mu_e}{\dd t} & = & M_e^{\ a}\left(\lambda_a^\alpha\mu_\alpha+X_a^0 \dot{n}_\mt{q} \right)\ ,\nonumber
\end{eqnarray}
where the coefficients read
\begin{eqnarray}
M_1^a & = & (\chi^{-1})_s^a-(\chi^{-1})_d^a \nonumber\\ 
M_2^a & = & (\chi^{-1})_u^a+(\chi^{-1})_e^a-(\chi^{-1})_d^a\\
M_{s,e}^{\ a} & = & (\chi^{-1})_{s,e}^{\ a}\ ,\nonumber
\end{eqnarray}
and
\begin{equation}
\lambda_a^\alpha=\left(
\begin{array}{cc}
0 & -(\lambda_2+\lambda_3)\\
\lambda_1 & \lambda_2\\
-\lambda_1 & \lambda_3\\
0 & -(\lambda_2+\lambda_3)
\end{array}
\right)\ .
\end{equation}
In terms of $\mu_1$ and $\mu_2$, the change in the pressure \eqref{eq:pressX} then becomes
\begin{equation}\label{eq:pressdef}
p=p_0+ \nq^0\left[ X_u^0 \delta \mu_2-(X_u^0+X_d^0) \delta\mu_1+\delta \mu_s-\frac{1}{3}\delta \mu_e\right]\ .
\end{equation}

At this point, we can reduce the problem to an algebraic system by expanding the chemical potentials in sine and cosine
\begin{eqnarray}
\mu_\alpha & = & \mu_\alpha^0+ \omega \Delta \nq\left[c_\alpha \cos(\omega t)+s_\alpha \sin(\omega t)\right]\ , \ \ \ \alpha=1,2 \ , \nonumber\\ 
\mu_s & = & \mu_s^0+ \omega \Delta \nq\left[c_s \cos(\omega t)+s_s \sin(\omega t)\right] \ ,\\
\mu_e & = & \mu_e^0+ \omega \Delta \nq\left[c_e \cos(\omega t)+s_e \sin(\omega t)\right]\ ,\nonumber
\end{eqnarray}
which lead to the set of equations ($\Gamma=1,2,s,e$ and $\beta=1,2$)
\begin{eqnarray}
(M_\Gamma\cdot \lambda^\beta)c_\beta-\omega s_\Gamma & = & - (M_\Gamma\cdot X^0) \nonumber\\
(M_\Gamma\cdot \lambda^\beta)s_\beta+\omega c_\Gamma & = & 0\ .\label{eq:cSeqs}
\end{eqnarray}
Introducing the pressure \eqref{eq:pressdef} in \eqref{eq:diss2} with the solutions we have written, we obtain now
\begin{equation}
\vev{{\cal E}_\text{diss}}=-\frac{\omega^2}{2}\left(\frac{\Delta \nq}{\nq^0}\right)^2(\nq^0)^2\left[X_u^0 c_2-(X_u^0+X_d^0) c_1+c_s-\frac{1}{3}c_e\right]\ ,
\end{equation}
and comparing with \eqref{eq:dissen}, we find a simple expression for the bulk viscosity,
\begin{equation}\label{eq:bulkvis}
\zeta=(\nq^0)^2\left[X_u^0 c_2-(X_u^0+X_d^0) c_1+c_s-\frac{1}{3}c_e\right]\ .
\end{equation}
Defining finally a matrix with components 
\begin{equation}\label{eq:calM}
{\cal M}_\alpha^{\ \beta}=\omega^2\delta_\alpha^\beta+(M_\alpha\cdot \lambda^\gamma)(M_\gamma \cdot \lambda^\beta)\ ,
\end{equation}
the solution for the cosine coefficients in \eqref{eq:cSeqs} becomes
\begin{eqnarray}
c_\alpha & = & -({\cal M}^{-1})_\alpha^{\ \beta} (M_\beta\cdot \lambda^\gamma) (M_\gamma \cdot X^0)\ , \ \ \ \alpha=1,2 \nonumber \\
c_{s,e} & = & -\frac{1}{\omega^2}(M_{s,e}\cdot \lambda^\beta)\left[ (M_\beta \cdot X^0)+(M_\beta \cdot \lambda^\gamma)c_\gamma\right]\ .\label{eq:Ccoefs}
\end{eqnarray}

\subsection{A.3: Non-leptonic limit}

To conclude the derivation of the bulk viscosity formulas, let us finally briefly study how the result simplifies is the leptonic processes are neglected by setting $\lambda_2=\lambda_3=0$ and only the non-leptonic processes $u+d\longleftrightarrow u+s$ retained. 
Using the symmetry of the susceptibility matrix, we now define the coefficients
\begin{eqnarray}
A_1 & = & \left[(\chi^{-1})_s^a-(\chi^{-1})_d^a\right]n_a^0 \nonumber\\
C_1 & = & M_1^s-M_1^d=(\chi^{-1})_d^d+(\chi^{-1})_s^s-2 (\chi^{-1})_d^s \\
D_1 & = & M_2^s-M_2^d=(\chi^{-1})_d^d-(\chi^{-1})_d^s+(\chi^{-1})_u^s+(\chi^{-1})_s^e-(\chi^{-1})_u^d-(\chi^{-1})_d^e\ ,\nonumber
\label{eqn:a1c1d1}
\end{eqnarray}
whereby the matrix ${\cal M}$ in \eqref{eq:calM} simplifies to
\begin{equation}
{\cal M}=\left( 
\begin{array}{cc}
\omega^2+C_1^2\lambda_1^2 & 0\\
 C_1 D_1 \lambda_1^2 & \omega^2
\end{array}
\right)\ .
\end{equation}
Solving for the coefficients in \eqref{eq:Ccoefs} and introducing the result in the bulk viscosity formula \eqref{eq:bulkvis}, we obtain\footnote{This formula relies on the constraints on the partial fractions in eq.~\eqref{eq:partfrac}.}
\begin{equation}
 \zeta=\frac{\lambda_1 A_1^2}{\omega^2+\lambda_1^2 C_1^2}\ ,
\end{equation}
i.e., eq.~\eqref{eq:zeta} from the main text. 
{\color{black}Further inverting the susceptibility matrix, we see that the coefficients $A_1$ and $C_1$ acquire the forms
\begin{eqnarray}
A_1 &=& H_p^{-1} \Big\lbrace n_s \left[\chi_{uu}\left(\chi_{dd}+\chi_{ds}\right)-\chi_{du}\left(\chi_{ud}+\chi_{us}\right)\right] -n_d \left[\chi_{uu}\left(\chi_{ss}+\chi_{sd}\right)-\chi_{su}\left(\chi_{ud}+\chi_{us}\right)\right]\nonumber \nonumber \\
&&+n_u \left[\chi_{ud}\left(\chi_{ss}+\chi_{sd}\right)-\chi_{us}\left(\chi_{dd}+\chi_{ds}\right)\right]\Big\rbrace \label{A1} \\
C_1 &=& H_{p}^{-1}\! \left[ (\chi_{ud}+\chi_{us})^2-\chi_{uu}(\chi_{dd}+2\chi_{ds}+\chi_{ss})\right]\!,\;\;\;\label{C1}
\end{eqnarray}
where $H_p=\det \chi_{ij}$ is the Hessian determinant of the pressure $p$. 
These results include the effects of off-diagonal susceptibilities, which to the best of our knowledge have not been included in the determination of $\zeta$ for unpaired quark matter before.}
If we finally assume that the susceptibilities are diagonal, $\chi_a^b=\chi_{aa}\delta_a^b$, the coefficients take the simple form
\begin{equation}\label{eq:diagonalcoeffs}
C_1=\frac{1}{\chi_{dd}}+\frac{1}{\chi_{ss}}\ ,\ \ 
A_1=\frac{n_d}{\chi_{dd}}-\frac{n_s}{\chi_{ss}}\ .
\end{equation}
in which case re-inverting the susceptibility matrix becomes trivial.

\section{B: Details on the V-QCD calculation}

Given the sensitivity of the bulk viscosity on the quark masses, we need to generalize the V-QCD model in a way that includes quark flavors with non-degenerate masses. 
While the flavor dependence is implicitly included in the model as defined in earlier literature~\cite{Jarvinen:2011qe,Alho:2013hsa}, the implications of flavor dependence have not been properly studied before. 
Here, we briefly review the computation of the QM equation of state and Nambu--Goldstone boson masses in the flavored case. 
We stress, however, that the flavor-dependent V-QCD extends the flavor-independent case in a rather simple fashion: to the most part, one just needs to sum over the contributions from each flavor.

The fields of the holographic dictionary relevant for the present setup are the following:
\begin{itemize}
    \item The dilaton $\phi$, which is a scalar field dual to ${\cal F}^2$, where ${\cal F}$ is the field strength operator of the gluon field. 
    The source corresponding to this operator is the gauge coupling of QCD.
    \item The tachyon $T^{ij}$, where $i,j=1\ldots \Nf$ are flavor indices, which is dual to the operator $\bar \psi^i \psi^j$ with $\psi^i$ the $i$-quark field in QCD. 
    The source corresponding to this operator is the quark mass matrix.
    \item The (vectorial) bulk gauge fields $A_{\mu}^{ij}$, which are dual to the currents $\bar \psi^i\gamma_\mu\psi^j$. 
    The sources for the temporal components include the chemical potentials of different quark flavors.
\end{itemize}
The relevant terms in the model action are on the other hand given by
\begin{equation}
    S_\mathrm{V-QCD} = S_\mt{g} + S_\mt{f} \ ,
\end{equation}
where $S_\mt{g}$ is the action of the improved holographic QCD (IHQCD) model~\cite{Gursoy:2007cb,Gursoy:2007er}. 
It describes the gluon sector of the theory via five-dimensional Einstein-dilaton gravity
\begin{equation}
    S_\mt{g} = \Mp^3 \Nc^2 \int \dd^{5} x\sqrt{-g}\left[R-\frac{4}{3} g^{M N} \partial_{M} \phi \partial_{N} \phi+\Vg(\phi)\right] \ ,
\end{equation}
where the Planck mass $\Mp$ and the potential $\Vg$ need to be determined by comparing to QCD data~\cite{Gursoy:2009jd}. 
The flavor action on the other hand takes the form of a generalized Dirac--Born--Infeld action~\cite{Bigazzi:2005md,Casero:2007ae}
\begin{align}
S_\mt{f}&=-\Mp^3 \Nc \int \dd^5x\,\mathrm{Tr}\left[\Vf(\phi,T)\sqrt{-\det (g_{MN}  + w(\phi) F_{MN}+\kappa(\phi)\partial_M T\partial_N T)}\right]  &\\
 & = -\Mp^3 \Nc \sum_{i=1}^{\Nf} \int \dd^5x\,\Vf(\phi,\tau_i)\sqrt{-\det (g_{MN}  + w(\phi) F_{MN}^{(i)}+\kappa(\phi)\partial_M \tau_i\partial_N \tau_i)} \ ,
\end{align}
where we assume that all flavor-dependent fields are diagonal: $T^{ij}=\tau_i\delta^{ij}$ and $(A_M)^{ij} = A^{(i)}_M \delta^{ij}$. 
Here, the capital Latin indices run over all five dimensions whereas the Greek indices denote the Lorentz indices in four dimensions.

The potentials $\Vg$, $\Vf$, $\kappa$, and $w$ in the above actions can be determined by comparing to QCD data~\cite{Gursoy:2009jd,Jokela:2018ers,Ishii:2019gta,Amorim:2021gat,Jarvinen:2022gcc}. 
In this article, we use the potential sets 5b, 7a, and 8b fitted to lattice thermodynamics in~\cite{Jokela:2018ers,Ishii:2019gta}. 
We have also checked that the potentials constructed in~\cite{Jarvinen:2022gcc}, which additionally use data for meson masses and decay constants, produce similar results.

We use the metric ansatz
\begin{equation} \label{metric}
\mathrm{d} s^{2}=e^{2A(r)}\left[-f(r) \mathrm{d} t^{2}+\mathrm{d} \mathbf{x}^{2}+\frac{\mathrm{d} r^{2}}{f(r)}\right], \quad f(0)=1 \ ,
\end{equation}
with the AdS$_5$ boundary located at $r=0$. 
As we are interested in the equilibrium thermodynamics of homogeneous phases, the metric as well as the background values for the gauge fields and scalars are assumed to be functions of $r$ only. 
We only consider black-hole backgrounds, which have a horizon at some $r=\rh$ where the blackening factor vanishes, $f(\rh)=0$. 
The temperature and the entropy density are then given by the surface gravity and area of the horizon:
\begin{equation} \label{eq:BHthermo}
    T = \frac{1}{4\pi}\left|f'(\rh)\right|\ , \qquad s = 4\pi \Mp^3 \Nc^2 e^{3 A(\rh)} \ . 
\end{equation}

We use the standard radial gauge $A_r^{(i)}=0$. 
In order to study the model at nonzero quark chemical potentials, we turn on temporal components $\Phi_i \equiv A_t^{(i)}$ for the gauge fields. 
The variables $\Phi_i$ are then cyclic: the action only depends on them via the derivatives $\Phi_i' \equiv \partial_r \Phi_i$. 
Consequently, the gauge fields can be integrated out leading to \cite{Alho:2013hsa}
\begin{equation}
    \hat n_i \equiv \frac{n_i}{\Mp^3\Nc^2} = - \frac{e^{A}   \Vf(\phi,\tau_i)w(\phi)^2\Phi_i'}{\Nc \sqrt{ 1+e^{-2A}f \kappa (\phi) \left(\tau_i'\right)^2-e^{-4 A}w(\phi)^2\left(\Phi_i'\right)^2 }}\ ,  
\end{equation}
where $n_i$ is the number of quarks with flavor $i$. 
The total quark number and baryon number become then 
\begin{equation}\label{eq:nqdef}
    \nq = \sum_{i=1}^{\Nf} n_i = \Mp^3 \Nc^2 \sum_{i=1}^{\Nf} \hat n_i \ , \qquad
    \nB = \frac{\nq}{\Nc} = \Mp^3 \Nc \sum_{i=1}^{\Nf} \hat n_i \ ,
\end{equation}
respectively.  
The gauge fields must vanish at the horizon, $\Phi_i(\rh)=0$, while the quark chemical potentials are given in terms of their boundary values,
\begin{equation}
    \mu_i = \left.\Phi_i\right|_{r=0} =  \Nc \hat n_i\int_0^{\rh} \dd r \frac{e^{-A}\sqrt{1+e^{-2A}f \kappa(\phi)\left(\tau_i'\right)^2}}{\Vf(\phi,\tau_i)w(\phi)^2\sqrt{1+\frac{\Nc^2\hat n_i^2}{e^{6A}\Vf(\phi,\tau_i)^2w(\phi)^2}}} \ .
\end{equation}

Finally, we also need to control the flavor-dependent quark masses. 
They are given as the sources of the tachyon fields, i.e., the coefficients $m_i$ in the boundary expansion~\cite{Jarvinen:2011qe}
\begin{equation} \label{eq:mqdef}
    \tau_i(r) = m_{i} r \ell \left[-\log(r\Lambda)\right]^{\gamma}\left[1+\mathcal{O}\left(\frac{1}{\log (r\Lambda)}\right)\right] \ ,
\end{equation}
where the AdS radius $\ell$ and the energy scale $\Lambda$ are determined by the boundary behavior of the metric and the dilaton
\begin{equation}
    A(r) = \log \frac{\ell}{r} + \mathcal{O}\left(\frac{1}{\log (r\Lambda)}\right) \ , \qquad \phi(r) = -\log\left(-\log \left(r\Lambda\right)\right) + \log\frac{8}{9v_1} +\mathcal{O}\left(\frac{1}{\log (r\Lambda)}\right) \ .
\end{equation}
In these formulas $\gamma$, $\ell$, and $v_1$ are found explicitly in terms of the near-boundary expansions of the potentials $\Vg$, $\Vf$, and $\kappa$, but $\Lambda$ needs to be determined numerically~\cite{Jarvinen:2011qe} and is typically around $\Lambda_\mathrm{QCD}$.  
Note that $m_i$ do not correspond to the physical quark masses but differ from those defined, e.g., in chiral perturbation theory by multiplicative constants. 
After introducing the numerical methods we employ in the V-QCD calculation below, we will briefly discuss the determination of the $m_s$ parameter by fitting the masses of pseudoscalar mesons.

\subsection{B.1: Numerical method}

In order to solve the backgrounds numerically, we first fix $\Nc=\Nf=3$ and $m_u=m_d=0$ and choose as the potentials the sets 5b, 7a, and 8b from~\cite{Jokela:2018ers,Ishii:2019gta}. 
As we are studying the deconfined phase without spontaneous chiral symmetry breaking, the fields $\tau_u$ and $\tau_d$ vanish identically. 
We then proceed to solving the Einstein equations and the equation for $\tau_s$ numerically by shooting from the horizon towards the boundary. 
Since the entropy in~\eqref{eq:BHthermo} is directly given in term of horizon quantities, it turns out that the natural input charges are the ratios
\begin{equation}
    \tilde n_i = \frac{4\pi n_i}{s} = \frac{\hat n_i}{e^{3 A(\rh)}} \ ;
\end{equation}
see~\cite{Alho:2013hsa} for details. 
The input values to the computation are then the horizon values of the scalars $\phi_h = \phi(\rh)$, 
$\tau_{sh}=\tau_s(\rh)$, and the three charges $\tilde n_u$, $\tilde n_d$, and $\tilde n_s$. 
After the solution has been found, we can compute $m_s$, $s$, $T$, $\mu_u$, $\mu_d$, and $\mu_s$ as functions of the input parameters using the formulas given above. {\color{black}The horizon value of $e^\phi$ can be identified as  the 't Hooft coupling in this setup~\cite{Gursoy:2007cb}, i.e., $\alpha_s = e^{\phi_h}/(12\pi)$ for $N_c=3$.}

The value of $\tau_{sh}$ is determined for each solution by fitting the strange quark mass to the pseudoscalar meson masses as we discuss below. 
As above, the relations between quark chemical potentials are solved by requiring charge neutrality and $\beta$-equilibrium after adding a free electron gas on top of the strongly coupled model. 
The susceptibilities $\chi_{ij}$ are then found as numerical derivatives of the various quark number densities with respect to the chemical potentials. 
Here, we use a brute force method where we vary all input parameters except for $\tau_{sh}$ on a small four-dimensional grid around a chosen point, and obtain the partial derivatives by fitting the results.

\subsection{B.2: Determining the strange quark mass parameter}

Finally, let us briefly discuss how the strange quark mass parameter $m_s$ is determined by fitting the masses of the pseudo-Goldstone bosons in QCD. 
For this we need to consider the ansatz
\begin{equation}
T = \mathrm{diag}(\tau_u,\tau_d,\tau_s)\,e^{i\pi_a t_a} \ ,    
\end{equation}
where the phases $\pi_a$ are small fluctuations and $t_a$ are the generators of SU(3). 
The fluctuation analysis is then a generalization of the flavor-independent analysis performed in~\cite{Arean:2012mq,Arean:2013tja} (To be precise, one also needs to fluctuate the gauge fields, the longitudinal components of which mix with the pseudoscalars). 
This generalization is largely straightforward, but there is a small complication: when the masses of different quark flavors are unequal, the background and the fluctuations involve matrices in flavor space that do not necessarily commute. 
The general prescription of the tachyonic Dirac--Born--Infeld actions for such cases is not known, but for the purposes of this article it is enough to adopt a simple prescription: we arrange the background matrices (functions of the background tachyon field) in the fluctuation action to the left and the fluctuation wave functions (combinations of $\pi_at_a$) to the right in the single trace appearing in the action for the fluctuations. 
We focus on the case of equal $u$ and $d$ quark masses, in which case $\tau_u=\tau_d \equiv \tau_{ud}$, and the standard pion, kaon, and $\eta$ states in $\pi_a t_a$ form a diagonal basis for the fluctuations. 
In this case, the ambiguity in defining the fluctuations will only affect the kaons, as pions and the $\eta$ commute with the background. 

\addtolength{\tabcolsep}{2pt}    
\begin{table}[]
    \centering
    \begin{tabular}{crrrrr}
        \toprule
         Potentials & $m_{ud}/\Lambda$ & $m_{s}/\Lambda$ & $m_\pi$[MeV] & $m_K$[MeV] & $m_\eta$[MeV] \\
         \midrule
         5b & 0.02809 & 0.6682 & 135.3 & 505.1 & 540.4  \\
         7a & 0.02479 & 0.5914 & 135.1 & 505.1 & 540.1 \\
         8b & 0.01988 & 0.5279 & 135.2 & 507.2 & 538.7 \\
        \midrule
         Experiment & & & ($\pi^0$) 135.0 & ($K^0$) 497.6  & 547.9  \\
         &  & & ($\pi^\pm$) 139.6 &  ($K^\pm$) 493.7 & \\ 
         \bottomrule
    \end{tabular}
    \caption{Fit results for the quark and meson masses in the flavor-dependent holographic model.}
    \label{tab:massfitvalues}
\end{table}
\addtolength{\tabcolsep}{-2pt}    

To proceed, we write the usual plane wave Ansatz for the fluctuations,
\begin{equation}
    \pi_a(r,x_\mu) = \phi_a(r) e^{iq^\mu x_\mu} \ , 
\end{equation}
arriving at the following fluctuation equations for the wave functions
\begin{equation} \label{eq:flucts}
    e^{2A}\kappa(\phi)\left[\sum_i c_{ia}\Vf(\phi,\tau_i)\tau_i^2G_i^{-1}\right]\frac{\dd}{\dd r}\left[
    \frac{\psi_a'}{e^{3A}\kappa(\phi)\sum_i c_{ia}\Vf(\phi,\tau_i)\tau_i^2G_i}\right] = \frac{4 e^A \kappa(\phi)}{w(\phi)^2}\frac{\sum_ic_{ia}\Vf(\phi,\tau_i)\tau_i^2G_i^{-1}}{\sum_ic_{ia}\Vf(\phi,\tau_i)G_i^{-1}}\psi_a +e^{-A}q^2\psi_a  \ ,
\end{equation}
where there is no sum over $a$, and
\begin{equation}
    G_i = \sqrt{1+e^{-2A}\kappa(\phi)\left(\tau_i'\right)^2} \ , \qquad \psi_a = e^{3A}\kappa(\phi)\left[\sum_ic_{ia}\Vf(\phi,\tau_i)\tau_i^2G_i^{-1}\right]\phi'_a \ .
\end{equation}
The index $i$ takes two values, $ud$ and $s$, and the coefficients $c_{ia}$ are determined such that
\begin{equation}
    \mathrm{Tr}\left[\mathrm{diag}(\tau_{ud},\tau_{ud},\tau_s) t_a^2\right] = c_{uda}\tau_{ud} + c_{sa}\tau_s \ . 
\end{equation}
If we adopt a normalization $\mathrm{Tr}\left[t_at_b\right] = \delta_{ab}/2$, the coefficients are given by
\begin{equation}
    c_{ud\pi} = \frac{1}{2} \ , \quad c_{s\pi} = 0 \ ; \qquad c_{udK} = \frac{1}{4} \ ,\quad  c_{sK} =  \frac{1}{4} \ ; \qquad \qquad c_{ud\eta} = \frac{1}{6} \ ,\quad  c_{s\eta} =  \frac{1}{3}\ .
\end{equation}

In order to fit the meson masses we construct the vacuum, i.e.,~the zero-temperature solution of V-QCD (see~\cite{Jarvinen:2011qe}) at finite quark masses and search for normalizable solutions to~\eqref{eq:flucts} with lowest mass $-q^2$ in each sector. 
We vary the quark masses determined through~\eqref{eq:mqdef} and search for the best fit to the experimental values of the pion, kaon, and $\eta$ masses (cf.~Table~\ref{tab:massfitvalues}) finding agreement with the experimental values within a few percent. 
In the analysis of the bulk viscosity, we set $m_{ud}$ to zero for simplicity, and use the values of $m_s$ from this table. 

As we are working in the deconfined phase at finite temperature and density, it would be better to fit the strange quark mass to some observables in this phase rather than to the masses of pseudoscalar mesons. 
Indeed, it has been found that simultaneously fitting the model using data from both the confined phase (such as meson masses) and the deconfined phase (such as lattice results for the EoS at zero density and high temperature) is challenging and easily leads to tensions in the fit parameters~\cite{Jarvinen:2022gcc}. 
In the deconfined phase, there is lattice data for quark number susceptibilities at high temperatures and small densities, showing strong dependence on the strange quark mass~\cite{Borsanyi:2011sw}. 
However, fitting the strange-quark-mass dependence using this data in combination with the flavored version of the V-QCD model described above does not lead to a good fit. 
This is not surprising: The quark mass dependence of the flavored V-QCD model has not been compared in detail to such data in the literature so far, and there may be additional parameters that one must include in the holographic setup.

Nevertheless, we are able to make a general observation on the nature of the quark-mass corrections. 
The magnitude of the strange-quark-mass dependence of the V-QCD quark number susceptibilities significantly underestimates the effect seen in lattice data when using the value of the strange quark mass that is fitted to the kaon and $\eta$ masses in Table~\ref{tab:massfitvalues}. 
While the lattice data describes QCD only at low densities, this observation strongly suggests that the quark-mass dependence remains underestimated at high densities. 
This is in agreement with our results for the quantities that depend on the parameter $A_1$ --- vanishing at zero quark mass --- in fig.~2 of the main text: While the density dependence of the V-QCD result appears to be very similar to that of the pQCD result, the overall normalization of the former is clearly lower. 
Furthermore, the difference is so large that no smooth interpolation between the results is possible. 
We are planning to carry out a detailed analysis of the fit to the quark number susceptibilities in future work.

\section{C: Analyzing the behavior of the results}

To keep the figures in the main text as readable as possible, we have relegated a finer analysis of our results to this section. 
Below, we in turn investigate the limiting behavior of the bulk viscosity at high densities, compare our NNLO pQCD result with a lower-order ones, and study the validity of various approximations concerning, e.g., the inclusion of only diagonal susceptibilities or only a part of the full temperature dependence of the result.
{\color{black}At the end of the section, we also study the effect of the nFL correction to the electroweak rate, visible as the prefactor in eq.~(\ref{eq:lambda1}), on the bulk viscosity. This study gauges the stability of our results with respect to a future addition of currently unknown $O(\alpha_s)$ corrections to the electroweak rate.}

\begin{figure}
    \centering
    \includegraphics[width=0.45\textwidth]{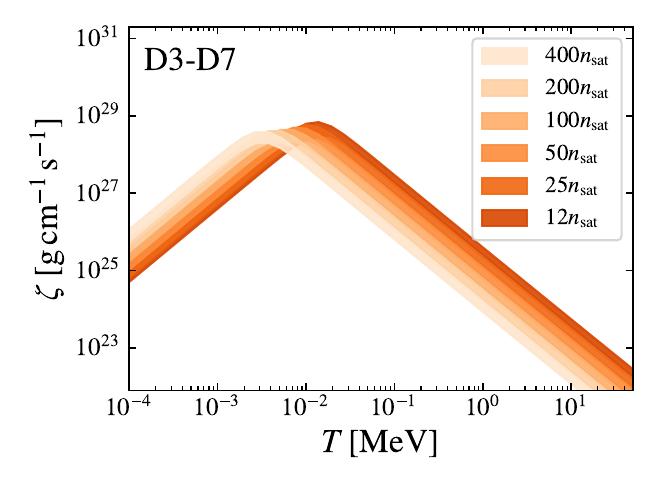}
    \includegraphics[width=0.45\textwidth]{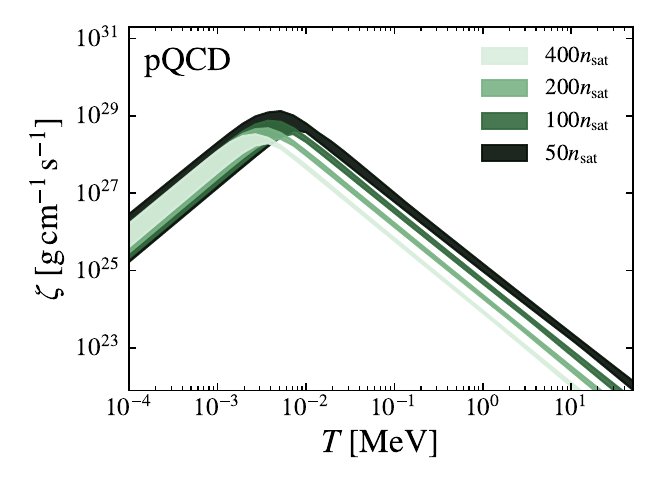}
    \caption{The bulk viscosity computed in the D3-D7 model and in pQCD as a function of $T$ over a large range of densities.}
    \label{fig:models_many_mu}
\end{figure}

\subsection{C.1: High-density limit and comparison with lower-order results}

\begin{figure}
    \centering
    \includegraphics[width=0.5\textwidth]{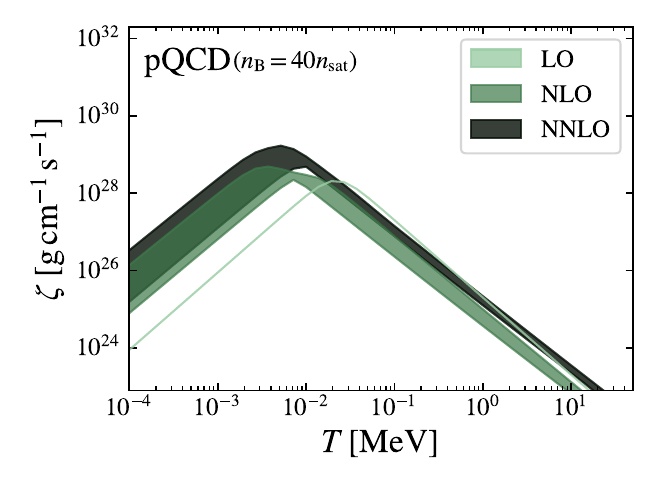}
    \caption{The bulk viscosity computed in pQCD at $40n_{\mathrm{sat}}$ using the LO (free quarks), NLO, and NNLO values for the pressure and its derivatives.}
    \label{fig:nlovsnnlo}
\end{figure}

Starting from the high-density limit, fig.~\ref{fig:models_many_mu} shows our D3-D7 and pQCD results as functions of temperature for a range of baryon densities above those displayed in fig.~\ref{fig:zeta_func_T_approx}. 
We observe a consistent pattern, where the peak value of the viscosity decreases and moves to smaller temperatures with increasing density, although in the D3-D7 case the peak value appears to ultimately saturate. 
Interestingly, below the peak temperature the value of the bulk viscosity increases as a function of density in both cases, but this behavior reverses after the peak. 
The density dependence is also seen to become stronger at higher temperatures, with the viscosity bands being quite tightly bundled together at lower values of $T$ especially in the pQCD result.

For pQCD, we are also able to compare our results with lower-order ones. 
This is done in fig.~\ref{fig:nlovsnnlo}, which shows our NNLO result together with a calculation where one only uses terms up to $\mathcal{O}(\alpha_s)$ in the weak-coupling expansion of the thermodynamic functions, including the mass correction of $\mathcal{O}(m_s^2)$ (``NLO''), as well as the leading-order (LO) result for non-interacting quarks. 
The NLO result is very similar to an existing NLO computation \cite{Sad:2007afd}, with small discrepancies arising due to the mass expansion scheme we use as well as the lack of a running coupling in \cite{Sad:2007afd}. 
{\color{black}One distinctive feature of these results is a clear increase in the renormalization-scale-related uncertainty at lower temperatures, which can be traced back to the logarithmically enhanced nFL corrections to the electroweak rate, visible in eq.~(\ref{eq:lambda1}).}

\subsection{C.2: The accuracy of various approximations}

\begin{figure}
    \centering
    \includegraphics[width=0.49\textwidth]{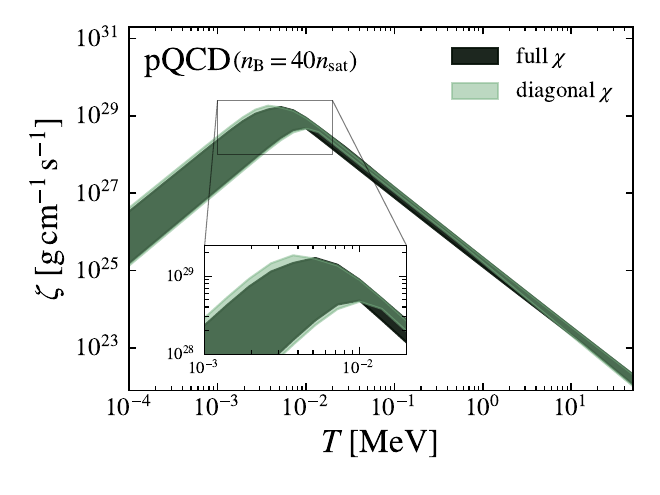}
    \includegraphics[width=0.49\textwidth]{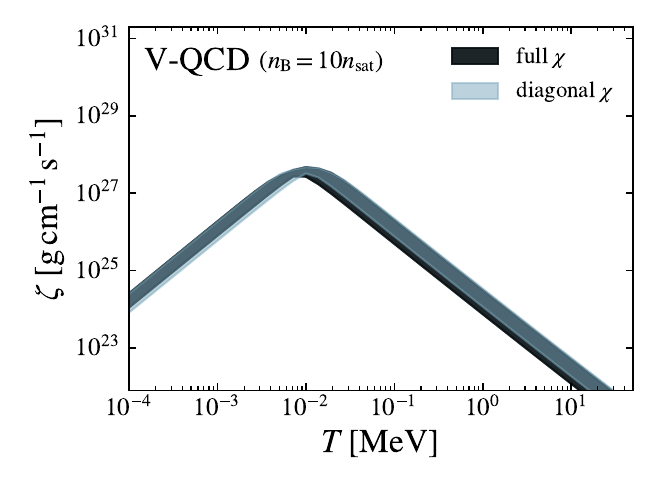}
    \caption{The bulk viscosity computed in pQCD (left) and in V-QCD (right), showing this time results obtained with the complete susceptibility matrix and its diagonal approximation. 
    In the left panel, the inset shows a zoomed-in bulk viscosity near the peak, emphasizing the small value of the difference.}
    \label{fig:diagonalvsnot}
\end{figure}

In the main text, the bulk viscosity is evaluated using the full susceptibility matrix, which is diagonal in the D3-D7 model due to the quarks being treated in the quenched approximation but contains off-diagonal elements in the pQCD and V-QCD calculations. 
As we can see from fig.~\ref{fig:diagonalvsnot}, which displays the latter two results in the diagonal approximation, the difference is, however, modest. 
For the pQCD calculation in particular, using the simplified eq.~(\ref{eq:diagonalcoeffs}) leads to a result that is numerically remarkably close to the full one. 
Nevertheless, we use the full expressions elsewhere in this letter, as the computational gain from using the diagonal approximation is quite small. 
In V-QCD, the difference of the two results is more significant, with the off-diagonal terms increasing the bulk viscosity below the peak and decreasing it at higher temperatures.

\begin{figure}
    \centering
    \includegraphics[width=0.78\textwidth]{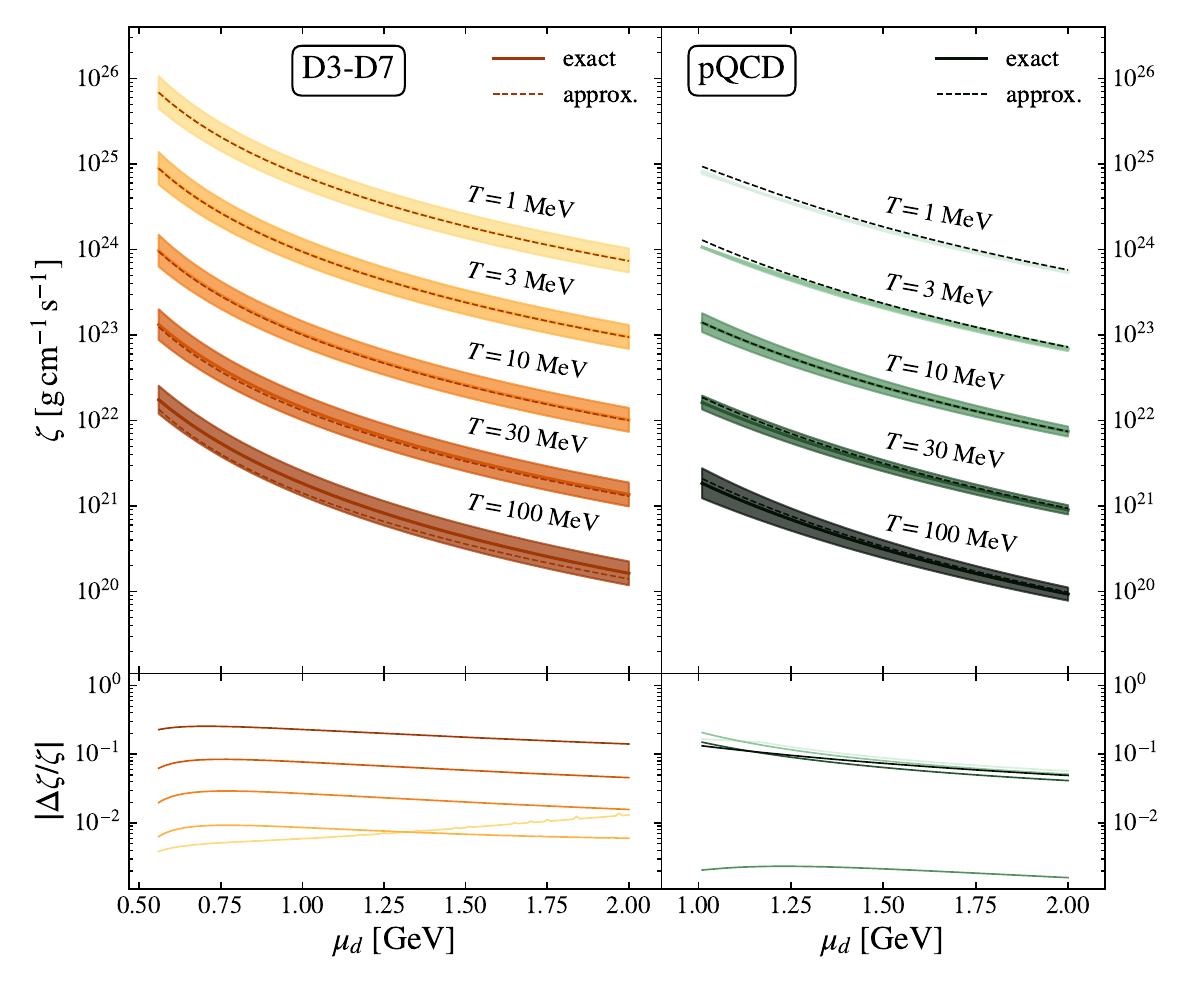}
    \caption{A comparison of our results for the bulk viscosity containing either the full or approximate temperature dependence. 
    The D3-D7 results are shown on the left and the pQCD ones on the right, with the full results always corresponding to solid lines with shaded uncertainty regions and the approximate ones to dashed lines. 
    The approximate results use the $T = 0$ values of $A_1$ and $C_1$, with all $T$ dependence originating from $\lambda_1$. 
    In this figure, the matching of the D3-D7 results to pQCD is performed at a fixed $\mu_d = 1$~GeV.}
    \label{fig:d3d7_exact_approx}
\end{figure}

Next, we briefly inspect the validity of our earlier claim that the temperature dependence of the bulk viscosity originates almost entirely from the electroweak rate in eq.~(\ref{eq:lambda1}), while $\zeta$ is nearly independent of the $T$-dependence of the quark densities and susceptibilities. 
This comparison is performed in fig.~\ref{fig:d3d7_exact_approx}, from which we indeed see that the full and approximate results are accurate up to $\mathcal{O}(10\%)$ corrections both in the D3-D7 and pQCD calculations for all $T \leq 100$~MeV. 
As argued before, this implies that we may evaluate the pressure and its derivatives at a fixed (small) temperature $T$, varying only the chemical potentials.

\begin{figure}
    \centering
    \includegraphics[width=0.78\textwidth]{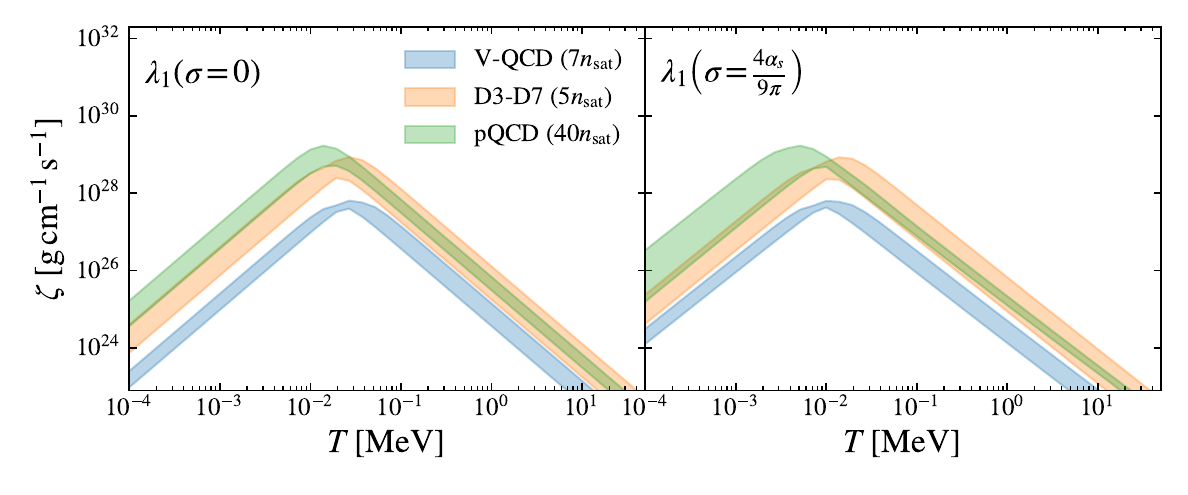}
    \caption{{\color{black}The effect of the nFL correction on the pQCD, D3-D7, and V-QCD results for the bulk viscosity, all evaluated at the lowest density where the results are available. The parameter $\sigma$ refers here to eq.~(\ref{eq:lambda1}), so that $\sigma=0$  corresponds to a result evaluated with the LO electroweak rate $\lambda_1$ and $\sigma=4\alpha_s/(9\pi)$ to a result where the nFL correction from \cite{Schwenzer:2012ga} has been included.}}
    \label{fig:nFLeffect}
\end{figure}

{\color{black}Finally, let us briefly look into the stability of our results with respect to potentially sizable QCD corrections to the electroweak rate $\lambda_1$. 
This is most straightforwardly achieved by comparing bulk-viscosity results evaluated with and without the nFL correction to eq.~(\ref{eq:lambda1}).
Precisely this is done in fig.~\ref{fig:nFLeffect}, where we observe, in accordance with the discussion of main text around fig.~\ref{fig:zeta_breakdown_3}, that the peak value and shape of the results stay intact, but the value of the peak temperature shifts due to the change in $\lambda_1.$ 
While the qualitative contrast between our QM results and the confined-phase curves in fig.~\ref{fig:zeta_func_T_approx} is left unchanged, this clearly highlights the importance of determining the remaining $O(\alpha_s)$ corrections to electroweak rate.}

\end{document}